\documentclass[pra,superscriptaddress,amsmath,amssymb,twocolumn]{revtex4}

\usepackage{graphicx}
\usepackage{dsfont}
\usepackage{color}

\begin{document}
\title{Quantum entanglement and extractable work for Gaussian states}

\author{Jaewon Lee}
\affiliation{Department of Physics, Kyungpook National University, Daegu 41566, Korea}
\author{Changsuk Noh}
\email{cnoh@knu.ac.kr}
\affiliation{Department of Physics, Kyungpook National University, Daegu 41566, Korea}
\author{Kabgyun Jeong}
\affiliation{Research Institute of Mathematics, Seoul National University, Seoul 08826, Korea}
\affiliation{School of Computational Sciences, Korea Institute for Advanced Study, Seoul, 02455, Korea}
\author{Hyunchul Nha}
\email{hyunchul.nha@qatar.tamu.edu}
\affiliation{Department of Physics, Texas A \& M University at Qatar, POBox 23874, Doha, Qatar}
\date{\today}

\begin{abstract}
The study of quantum thermodynamics aims to elucidate the role played by quantum principles in the emergent features of quantum thermodynamic processes. Specifically, it is of fundamental importance to understand how quantum correlation among different parties enables thermodynamic features distinguishable from those arising in classical thermodynamics. In this work, we investigate the relation between extractable work and quantum correlations for two-mode Gaussian states. We examine the change in local energy occurring at one party due to a Gaussian measurement performed on the other in relation to the quantum correlations of two-mode states classified as separable, entangled, and steerable states. Our analysis reveals a clear quantitative difference in the extractable work, depending on the class of states to which the two-mode state belongs.   
\end{abstract}

\maketitle

\section{Introduction}
Recent progress in experimental capabilities to prepare, manipulate, and measure quantum states with high precision has attracted much interest in exploring thermodynamics beyond the classical domain \cite{Mahler}. Numerous studies have been conducted to elucidate how quantum principles influence the emergent thermodynamic phenomena that can be distinguished from those predicted in classical theories. For instance, the celebrated fluctuation theorem (FT) \cite{FT1,FT2}, which describes the statistics of stochastic work or the entropy production via an equality form in non-equilibrium dynamics, has been  extended to quantum fluctuation theorems in various settings \cite{QFT1,QFT2,QFT3,QFT4,QFT5,QFT6,QFT7,QFT8}. These extensions include FT for quantum maps \cite{QCFT,QCFT1}, quantum coherence \cite{QCFT2,QCFT3,QCFT4} and quantum correlations \cite{QCF6,QCF7}.

\begin{figure}[!ht]
    \centering
    \includegraphics[width=0.9\columnwidth]{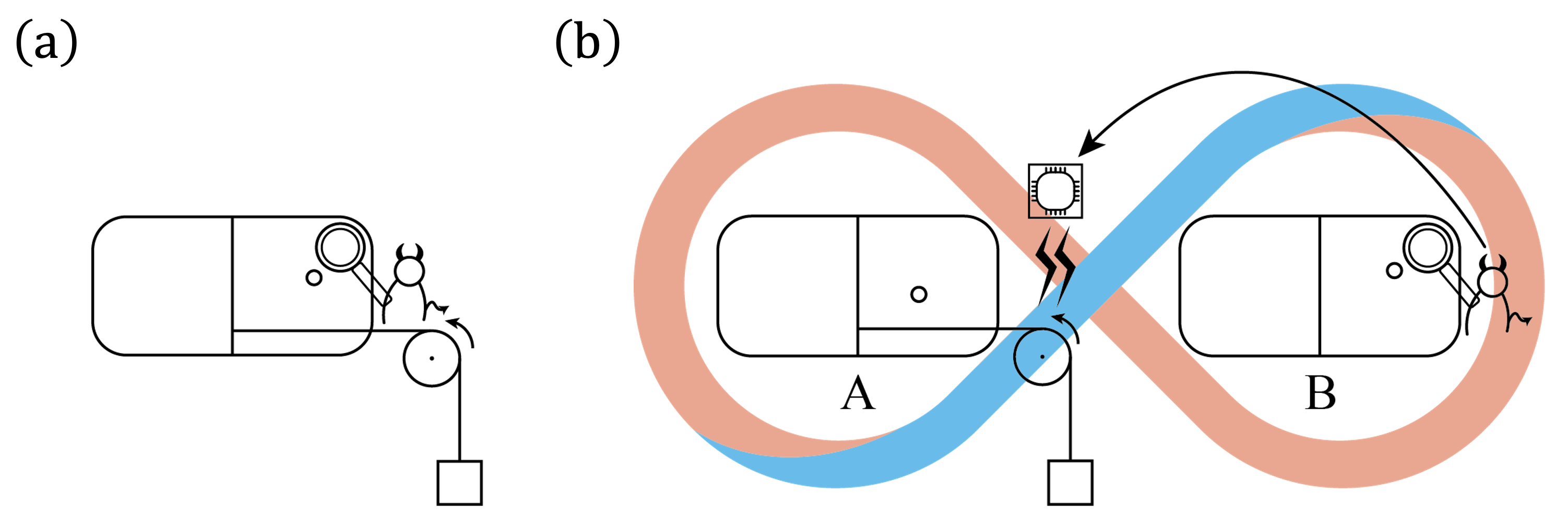}
    \caption{(a) Schematic illustration of the Szilard engine. Maxwell's demon observes the location of a particle and uses the information to extract work. (b) In the case of an entangled bipartite system, the demon may perform a measurement at the remote party B and use the information to extract work at the location of party A.}
    \label{fig:szilard}
\end{figure}

It is of fundamental importance to identify quantum features responsible for differences between classical and quantum thermodynamic processes. We here aim to find such a feature by investigating an extractable work in a context similar to Szilard's engine \cite{Sz1,Sz2}. In the Szilard engine, the so-called Maxwell's demon obtains information on the location of a particle in a box by performing a measurement and uses the information to extract work (see Fig.~\ref{fig:szilard}(a)). 
In this process, the extracted work is quantified by the change in the entropy of the system as $W=k_BTS(\rho_i)-k_BTS(\rho_f)$, which highlights the role played by information in a heat engine ($k_B$: the Boltzmann constant, $T$: temperature, $\rho_{i,f}$: initial \& final states). It raises an interesting question of whether quantum information obtained via  quantum measurement can enhance extracted work beyond what is achievable with classical information. In particular, when a quantum bipartite system is entangled, the measurement by the demon can be made on the remote entangled party (see Fig.~\ref{fig:szilard}(b)). 
In this scenario, the amount of locally extracted work can depend on the degree of quantum correlation between the two parties. If a classical bound on this work is found that can be achieved without entanglement, demonstrating enhanced work may then serve as a quantum signature of the information heat engine. 

In this paper, we investigate the amount of locally extractable work for a general two-mode Gaussian state in relation to its quantum correlation. 
As work is not an observable but a representation of a physical process \cite{QW1,QW2}, a number of different approaches are possible for quantifying the amount of work \cite{QW3,QW4,QW5,QW6}. For instance, Ref.~\cite{Mauro} defines work in terms of the R\'enyi-2 entropy in order to investigate the relationship between work extraction and correlations (separable vs entanglement) in the Gaussian framework, i.e., Gaussian states and Gaussian measurements. In contrast, our approach deals with the change in local energy, which can be more directly interpreted as the extractable work. In principle, such change in local energy can be converted into useful work by coupling the system to an ancillary work-storage system. An energy-conserving unitary operation between the working medium and the work-storage system transfers the reduced energy in the former to the energy gain in the latter \cite{Alhambra,Strunz}. 

Different scenarios for work extraction may have a bearing on quantum correlation in different contexts. For instance, Ref.~\cite{discord} investigates work extraction through a cyclic unitary operation applied to bipartite quantum states. The optimized work in this scenario, known as ergotropy, is quantified by the energy difference between the initial and final states and is bounded by the quantum mutual information, which consists of quantum correlations (quantum discord) and classical correlations (the Holevo bound). In this case, the unitary operation may include a nonlocal operation as it brings an initially correlated state to a passive state close to a product state. 
In contrast, we aim to elucidate the relationship between quantum entanglement and locally extractable work.
Therefore, our work extraction scenario involves an LOCC protocol suitable for characterizing quantum entanglement.
In doing so, we demonstrate a strong overall relationship between the amount of locally extractable work and the degree of quantum correlation, as reflected in the hierarchy of separable, entangled, and steerable states \cite{Wiseman}.

\section{Gaussian states and measurements}
Gaussian states and measurements are special classes of states and measurements in continuous-variable quantum systems that are relatively easy to prepare and perform. This section introduces the formalism for handling such systems and sets conventions and notations along the way.

\subsection{Gaussian states}
A general Gaussian state can be characterized by its first moments (average amplitudes) and second moments (covariances). Let us represent a collection of canonical quantum operators as   
    $\bold{\hat{R}}=(\hat{x}_{1},\hat{p}_{1},...,\hat{x}_{n},\hat{p}_{n})^{\mathrm{T}}, $ with their averages $\bold{\overline{R}}=\mathrm{Tr}[\rho\bold{\hat{R}}] \equiv \langle  \bold{\hat{R}} \rangle$ for a given state $\rho$. Here $\hat{x}_{i} \equiv \frac{1}{\sqrt{2}}(\hat{a}_{i}+\hat{a}^{\dagger}_{i})$ and $\hat{p}_{i} \equiv \frac{1}{\sqrt{2}i}(\hat{a}_{i}-\hat{a}^{\dagger}_{i})$ are the canonical observables for mode $i$ satisfying $[{\hat x}_i,{\hat p}_i]=i$, or equivalently $[\hat{a}_{i},\hat{a}^{\dagger}_{i}]=1$.

 The elements of the covariance matrix for $\rho$ are given by $\Gamma_{ij}\equiv\langle \Delta R_i \Delta R_j+\Delta R_j\Delta R_i\rangle$ where $\Delta O\equiv\hat O-\langle \hat O\rangle$. 
 For the case of two-mode states shared by Alice ($A$) and Bob ($B$), with $\bold{\hat{R}}=({\hat x}_A,{\hat p}_A,{\hat x}_B,{\hat p}_B)$, an arbitrary covariance matrix can be put into the standard form \cite{Duan} given by  
\begin{eqnarray}
 \Gamma_{AB} \equiv 
        \begin{pmatrix}
        \Gamma_{A} &  C  \\
        C^T &  \Gamma_{B}  \\
        \end{pmatrix}
        =
        \begin{pmatrix}
        a & 0 & c_{1} & 0 \\
        0 & a & 0 & c_{2} \\
        c_{1} & 0 & b & 0 \\
        0 & c_{2} & 0 & b \\
        \end{pmatrix}. 
        \label{eqn:standard}
 \end{eqnarray}       
Operationally, any given two-mode Gaussian state can be transformed to a state with the above covariance matrix via local unitary operations, which do not alter the entanglement properties of the two-mode state at all. Thus, the above standard form can be regarded as representing a general two-mode Gaussian state in examining the role played by quantum correlations.

For a covariance matrix to represent a physical state, a general uncertainty relation must be satisfied. This can be addressed in terms of a symplectic matrix $\Omega\equiv\oplus_i\Omega_i$ where 
$\Omega_i=\begin{pmatrix}
        0 &  1  \\
        -1 &  0  \\
        \end{pmatrix}$, with the uncertainty relation given by
$\Gamma+i\Omega\ge0$ \cite{Simon1}.

\subsection{Gaussian measurements}
A measurement is called Gaussian if the projected state after the measurement is a Gaussian state. Thus, a general Gaussian measurement can also be represented by its first moments (measurement outcomes) and second moments. Here, we confine ourselves to two important classes of Gaussian measurements, namely homodyne and heterodyne measurements.
These two types of measurements are covered by a single (measurement) covariance matrix given by $\Gamma_{m}=
\begin{pmatrix}
\lambda &  0  \\
0 &  \lambda^ {-1}  \\
\end{pmatrix}$. 
In the homodyne case, measurements corresponding to the $\hat x$- and $\hat p$-quadratures are described by $\lambda=0$ and $\infty$, respectively. In contrast, the heterodyne case corresponds to a simultaneous measurement of both $\hat x$- and $\hat p$-quadratures, with added noise accounted for by  $\lambda=1$. Other values of $\lambda$ represent different Gaussian measurements not considered in this work.
\subsubsection{Alice's state after Bob's measurement}
Below, we will need Alice’s state conditioned on Bob’s measurement result. To understand how such a state can be described, let us consider a two-mode state with local displacements $\bold{\overline{r}}_{A}$ and $\bold{\overline{r}}_{B}$, and   the covariance matrix $\Gamma_{AB}$ given in Eq.~(\ref{eqn:standard}).
If Bob performs a Gaussian measurement described by the covariance matrix $\Gamma_m$ on his mode $B$ and obtains a measurement outcome $\bold{r}_{m}$, Alice's state reduces to one \cite{weed} with a displacement given by 
\begin{eqnarray}
        \bold{\overline{r}}_{A} \rightarrow \bold{\overline{r}}_{A}+C(\Gamma_{B}+\Gamma_{m})^{-1}(\bold{r}_{m}-\bold{\overline{r}}_{B})
    \label{eq:dis_chg}
\end{eqnarray}        
        and a covariance matrix given by 
 \begin{eqnarray} 
        \Gamma_{A}\rightarrow \Gamma_{A}-C(\Gamma_{B}+\Gamma_{m})^{-1}C^{\mathrm{T}}. 
        \label{eq:cov_chg}
\end{eqnarray}
Noticeably, the covariance matrix of Alice's reduced state is independent of Bob's measurement outcome $\bold{r}_{m}$.
     
\section{Classification of two-mode Gaussian states}
Two-mode Gaussian states can be classified according to the types of correlations they exhibit. We consider three classes of states--separable, entangled, and steerable states--organized by increasing correlation strength, which are fully characterized by the covariance matrix. 

\subsection{Separable versus Entangled}
     A state is called separable if it can be written as a mixture of product states, i.e. $\rho_{AB}=\sum_{i}p_{i}\rho^{i}_{A} \otimes \rho^{i}_{B}$, where $\rho^{i}_{A}$ and $\rho^{i}_{B}$ are the states of subsystems A and B, respectively, with $\Sigma_{i}p_{i}=1$. If a state cannot be written in such a form,  it is entangled. 
     
Firstly, a two-mode Gaussian state with a covariance matrix $\Gamma_{AB}$ must satisfy the uncertainty relation, which can be written as 
\begin{eqnarray}
\Gamma_{AB}+
        \begin{pmatrix}
        i\Omega_{A} &  0  \\
        0 & i\Omega_{B}  \\
        \end{pmatrix} \ge 0,
       \label{eqn:phy}
 \end{eqnarray}      
 where $\Omega_{A}$ and $\Omega_{B}$ are the symplectic matrices introduced earlier. If this condition is violated, the corresponding state is not a valid quantum state.
 
Similarly, a two-mode Gaussian state is known to be separable \cite{Simon2} if and only if
\begin{eqnarray}
       \tilde{\Gamma}_{AB}+
        \begin{pmatrix}
        i\Omega_{A} &  0  \\
        0 & i\Omega_{B}  \\
        \end{pmatrix} \ge 0,
        \label{eqn:sep}
\end{eqnarray}
where $\tilde{\Gamma}_{AB}=\Lambda\Gamma_{AB}\Lambda^T$, with $\Lambda\equiv{\rm diag}\{1,1,1,-1\}$, represents the covariance matrix of the partially-transposed state.  Therefore, if the covariance matrix satisfies Eq.~(\ref{eqn:phy}) but violates Eq.~(\ref{eqn:sep}), it represents an entangled state.

\subsection{Steerable versus nonsteerable}
Quantum entangled states can be further classified into steerable and non-steerable states, with quantum steering representing a stronger form of quantum correlation \cite{Wiseman}. The latter can be defined in terms of probability distributions for measurement outcomes as follows. If a joint probability distribution $P(a,b:A,B)$ to obtain outcomes $a$ and $b$ for observables $A$ and $B$ can be written as $P(a,b:A,B)=\sum_\lambda p(\lambda)P^Q(a|A,\lambda)P(b|B,\lambda)$ by introducing a hidden variable $\lambda$ and its probability distribution $p(\lambda)$, it is nonsteerable from Bob to Alice. If a given probability distribution cannot be written as such, the corresponding state is steerable.

In the above definition of nonsteerability, note that the probabilities considered at Alice's site are restricted to quantum-mechanically allowed ones, $P^Q(a|A,\lambda)$, whereas Bob's probabilities, $P(b|B,\lambda)$, have no such restriction. Therefore, steering is an asymmetric notion of quantum correlation and the conditions for Alice $\rightarrow$  Bob steering are different from those for Bob $\rightarrow$ Alice steering. For a two-mode Gaussian state, the nonsteerability condition for Bob  $\rightarrow$ Alice case can be written as \begin{eqnarray}
\Gamma_{AB}+
        \begin{pmatrix}
        i\Omega_{A} &  0  \\
        0 & 0  \\
        \end{pmatrix} \ge0.   
\label{eqn:nonst}
\end{eqnarray}
The violation of this inequality confirms Bob  $\rightarrow$ Alice steerability. The non-steering condition for Alice $\rightarrow$  Bob is similarly given by $\Gamma_{AB}+
        \begin{pmatrix}
        0 &  0  \\
        0 & i\Omega_{B}  \\
        \end{pmatrix} \ge0 $, violation of which confirms Alice  $\rightarrow$ Bob steerability. 
         
\section{Extractable work}
\begin{figure}[!ht]
    \centering
    \includegraphics[width=0.9\columnwidth]{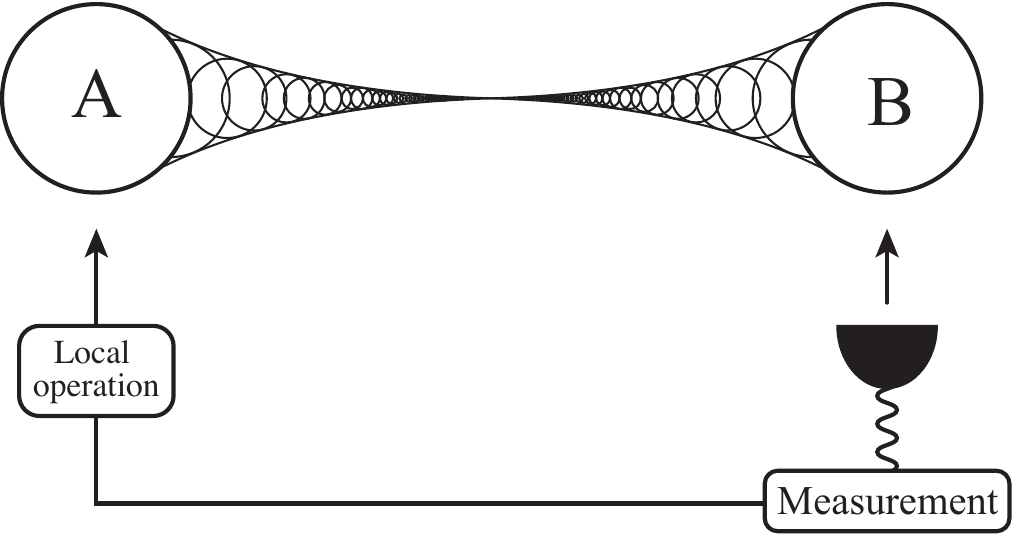}
    \caption{Schematic representation of the work extraction protocol. i) Bob performs a Gaussian measurement on his subsystem and sends the information to Alice. ii) Based on the measurement outcome, Alice performs Gaussian local operations to minimize the local energy. The amount of extractable work is defined as the difference in local energy between the initial and final states of Alice's subsystem. }
    \label{fig:protocol}
\end{figure}

We now turn to our proposed protocol, which uses information obtained from Bob's measurement to extract work at Alice's site. Bob's measurements are restricted to Gaussian measurements and Alice is only allowed to perform local Gaussian operations conditioned on Bob's measurement results (see Fig.~\ref{fig:protocol}). In this way, we identify the role of quantum correlations in work extraction. Without loss of generality, we consider the cases with zero initial displacements $\bold{\overline{r}}_{A}=\bold{\overline{r}}_{B}=0$, which can always be accomplished by {\it local} displacement operations, thus not affecting correlation structures at all. 

We define an extractable work as the difference in local energy between the initial state and the final state possessed by Alice, i.e. 
\begin{eqnarray}
    W = {\rm Tr} \left[a^\dag a\rho_A^i\right] -  {\rm Tr} \left[a^\dag a\rho_A^f\right].
\end{eqnarray}
This quantity can be easily calculated by noting that the local energy at Alice's site can generally be expressed as 
\begin{align}
\langle a^\dag a\rangle +\frac{1}{2} =\frac{1}{2}\left(\langle x^2\rangle+\langle p^2\rangle \right) =\frac{1}{2}\left(\langle x\rangle^2+\langle p\rangle^2+\Delta x^2+\Delta p^2 \right) =\frac{1}{2}\left(|\bold{\overline{r}}_{A}|^2+\frac{1}{2}{\rm Tr} \Gamma_A\right).
\label{eqn:energy}
\end{align}
In principle one can consider a work-storage system coupled to a working medium (Alice's mode in our case). If a total-energy conserving unitary operation is performed on those two systems, the reduction of local energy in the working medium then leads to the gain of energy in the work-storage system \cite{Alhambra}. How such an interaction can be practically implemented for continuous variable systems is an interesting topic, which will be dealt with elsewhere.

Starting with a two-mode state of zero displacements and the covariance matrix in Eq.~(\ref{eqn:standard}), the initial energy of Alice's mode is given by $\frac{1}{2}a$. To maximize the extractable work, we need to minimize both the displacement and the variances of the final state as can be seen from Eq.~(\ref{eqn:energy}). This leads us to the following protocols for work extraction.

\subsection{Protocol for homodyne detection}
In order to maximize the use of quantum correlation, we allow Bob to make measurements of complementary observables $\hat{x}$ and $\hat{p}$. The protocol for each type of measurement is as follows.

{\bf $X$-measurement protocol}:
(i) Bob performs a homodyne measurement of the $\hat x$-quadrature and obtains a certain outcome $x_m$. Using the measurement covariance matrix $\Gamma_m={\rm diag}\{0,\infty\}$, one finds that Alice's state after measurement is characterized by the displacement $\bold{\overline{r}}'_{A}=(\frac{c_1x_m}{b},0)$ and the covariance matrix $\Gamma'_A={\rm diag}\{a-\frac{c_1^2}{b},a\}$ from Eqs.~(\ref{eq:dis_chg}) and (\ref{eq:cov_chg}). (ii) Alice performs a local displacement along the $x$-quadrature by an amount $-\frac{c_1x_m}{b}$ to remove the measurement-induced displacement. (iii) Alice subsequently performs a local squeezing described by $\Gamma^{\rm final}_A=S\Gamma'_AS^T$,
where $S\equiv{\rm diag}\{s,1/s\}$ with $s^{-2}=\sqrt{1-\frac{c_1^2}{ab}}$. This completes the protocol for $X$-measurement, with the final energy given by $\frac{1}{2}a\sqrt{1-\frac{c_1^2}{ab}}$. 

The squeezing in step (iii) is designed to minimize the sum of the two variances: Consider an initial covariance matrix $\Gamma={\rm diag}\{\gamma_1,\gamma_2\}$. After applying a squeezing $S\equiv{\rm diag}\{s,1/s\}$, its trace $\gamma_1s^2+\gamma_2/s^2$ is minimized at $s^2=\sqrt{\gamma_2/\gamma_1}$. The effect of this squeezing is to make both of the variances identical. 

{\bf $P$-measurement protocol}:
(i) Bob performs a homodyne measurement of the $\hat p$-quadrature and obtains a certain outcome $p_m$. 
Using the measurement covariance matrix $\Gamma_m={\rm diag}\{\infty,0\}$, one finds that Alice's state after measurement is characterized by the displacement $\bold{\overline{r}}'_{A}=(0,\frac{c_2p_m}{b})$ and the covariance matrix $\Gamma'_A={\rm diag}\{a,a-\frac{c_2^2}{b}\}$ from Eqs.~(\ref{eq:dis_chg}) and (\ref{eq:cov_chg}). 
(ii) Alice performs a local displacement along the $p$-quadrature by an amount $-\frac{c_2p_m}{b}$ to remove the measurement-induced displacement. (iii) Alice subsequently performs a local squeezing 
described by $\Gamma^{\rm final}_A=S\Gamma'_AS^T$,
where $S\equiv{\rm diag}\{1/s,s\}$ with $s^{-2}=\sqrt{1-\frac{c_2^2}{ab}}$. This completes the protocol for $P$-measurement, with the final energy given by $\frac{1}{2}a\sqrt{1-\frac{c_2^2}{ab}}$. 

Combining the two protocols, the total extractable work on average is obtained as 
\begin{align}
W_{\rm hom}=\frac{a}{2}-\frac{a}{4}\left(\sqrt{1-\frac{c_1^2}{ab}}+\sqrt{1-\frac{c_2^2}{ab}}\right),   
\end{align}
which does not depend on the measurement outcomes.

\subsection{Protocol for heterodyne detection}
$\hat{x}$ and $\hat{p}$ can be measured simultaneously via heterodyne detection at the cost of added vacuum noise. This provides an alternative work-extraction protocol as follows.

(i) Bob performs a heterodyne measurement of the $\hat x$- and $\hat p$-quadratures simultaneously and obtains certain outcomes $x_m$ and $p_m$. 
With the covariance matrix $\Gamma_m={\rm diag}\{1,1\}$ of this Gaussian measurement, one finds that Alice's state after the measurement is characterized by the displacement $\bold{\overline{r}}'_{A}=(\frac{c_1x_m}{b+1},\frac{c_2p_m}{b+1})$ and the covariance matrix $\Gamma'_A={\rm diag}\{a-\frac{c_1^2}{b+1},a-\frac{c_2^2}{b+1}\}$ from Eqs.~(\ref{eq:dis_chg}) and (\ref{eq:cov_chg}). 
(ii) Alice performs a local displacement by an amount $\frac{c_1x_m}{b+1} +i \frac{c_2p_m}{b+1}$ to remove the measurement-induced displacement. (iii) Alice subsequently performs a local squeezing 
$\Gamma^{\rm final}_A=S\Gamma'_AS^T$,
where $S\equiv{\rm diag}\{1/s,s\}$ with $s^{-2}=\sqrt{\frac{a(b+1)-c_1^2}{a(b+1)-c_2^2}}$. This completes the protocol for heterodyne measurement, with the final energy given by 
$\frac{1}{2}\sqrt{a-\frac{c_1^2}{b+1}}\sqrt{a-\frac{c_2^2}{b+1}}$. The extractable work is then
\begin{align}
W_{\rm het} = \frac{a}{2} - \frac{a}{2}\sqrt{1-\frac{c_1^2}{a(b+1)}}\sqrt{1-\frac{c_2^2}{a(b+1)}},
\end{align}
which does not depend on the measurement outcomes.

\section{Results}
We aim to investigate the amount of extractable work for a general two-mode Gaussian state in relation to its quantum correlation. To achieve this, we divide the parameter space--spanned by the four variables  $a$, $b$, $c_1$ and $c_2$--according to the types of correlations and compare the resulting `phase diagram' with the extractable work under the homodyne and heterodyne protocols. Because of the difficulty in visualizing the four dimensional parameter space, we start by considering the simplest nontrivial case in which the local energies and the correlation strengths are equal, i.e., $a=b$ and $|c_1| = |c_2|$, and gradually increase the generality.

\subsection{$a=b$ and $c_1=-c_2$}
In this case, the local energies of the subsystems are equal and  the $X$- and $P$-correlations have the same strength. We only need to consider $c\equiv c_1=-c_2$, because only separable states are obtained for $c_1 = c_2$. The total extractable work in the homodyne protocol becomes $W_{\rm hom}=\left(a-\sqrt{a^2-c^2}\right)/2$, while that in the heterodyne protocol becomes $W_{\rm het}=c^2/\left(2a+2\right)$.

\begin{figure}[!ht]
\centering
\includegraphics[width=0.99\columnwidth]{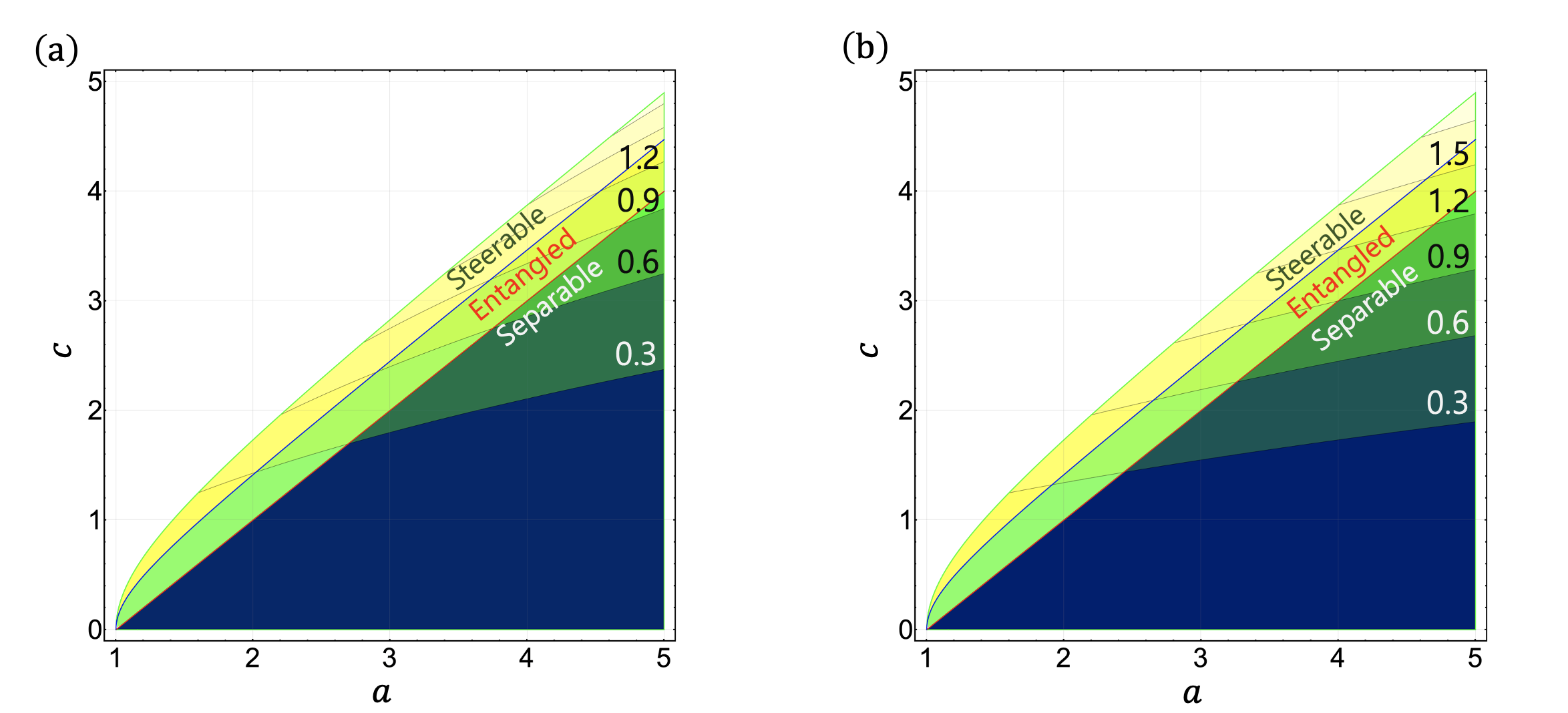}
\caption{Parameters regions for different types of correlated states overlaid with contour plots of the extractable work as functions of $a$ and $c$ under (a) homodyne measurement and (b) heterodyne measurement. Values in the contour denote the amount of extractable work. For a fixed value of $a$, the amount of extractable work increases with increasing correlation strength $c$. Furthermore, extractable work for different types of correlated states occupy distinct regions of the parameter space spanned by $a$ and $c$.}
\label{result:fig1}
\end{figure}

For this restricted class of states, the separability condition in Eq.~(\ref{eqn:sep}) becomes $c\le a-1$: The state is entangled if the correlation $c$ is greater than $a-1$. On the other hand, the nonsteerability condition in Eq.~(\ref{eqn:nonst}) becomes $c \le \sqrt{a^2-a}$, so the state is steerable if $c > \sqrt{a^2-a}$. Figure \ref{result:fig1} illustrates the parameter regions for separable, entangled and steerable states, respectively, separated by solid curves.
The same figure also displays the amount of extractable work, $W_{\rm hom}$ and  $W_{\rm het}$, represented by contour plots, where the extractable work increases within a given region as the color becomes lighter.  Different regions are color coded differently for clarity. The contour lines for $W_{\rm hom}$ and $W_{\rm het}$ do not coincide with the lines delineating the different types of correlated states due to the interplay between $a$ (initial local energy) and $c$ (correlation). A fair comparison, however, can be made for a fixed value of $a$. That is, for a given value of $a$, the work $W_{\rm hom}$ and $W_{\rm het}$ increases monotonically with the strength of correlation, $c$. In other words,  $W_{\rm hom}$ and $W_{\rm het}$ both obey the hierarchy $W^{\rm separable}< W^{\rm entangled}<W^{\rm steerable}$. A quick comparison shows that more work is extracted in the heterodyne protocol.

\subsection{$a=b$ and $c_1\neq c_2$ }

Next, let us relax the condition $|c_1| = |c_2|$ and consider the three-parameter family of states characterized by $a=b$, $c_1$ and $c_2$. In this case, the extractable work using the homodyne and heterodyne protocols reduce to
\begin{align}
&W_{\rm hom}=\frac{a}{2}-\frac{a}{4}\left(\sqrt{1-\frac{c_1^2}{a^2}}+\sqrt{1-\frac{c_2^2}{a^2}}\right)   , \\ 
&W_{\rm het} = \frac{a}{2} - \frac{a}{2}\sqrt{1-\frac{c_1^2}{a(a+1)}}\sqrt{1-\frac{c_2^2}{a(a+1)}}.
\end{align} 
We here focus only on a single quadrant of the phase space, occupied by $c_1 \le 0$ and $c_2 \ge 0$, because (i) the extractable work is symmetric with respect to the interchange $c_1 \leftrightarrow c_2$, and (ii) only separable states are obtained when the signs of $c_1$ and $c_2$ are equal.
Within the quadrant, the separability condition in Eq.~(\ref{eqn:sep}) becomes $c_2 \le \frac{c_1}{a^2 - c_1^2} + \sqrt{a^2 + \frac{a^2}{(c_1^2-a^2)^2} + \frac{2a^2}{c_1^2-a^2}}$, and the nonsteerability condition in Eq.~(\ref{eqn:nonst}) becomes $c_2 \le \sqrt{a \left( a + \frac{a}{c_1^2-a^2} \right)}$. 

\begin{figure}[!h]
    \centering
    \includegraphics[width=0.95\columnwidth]{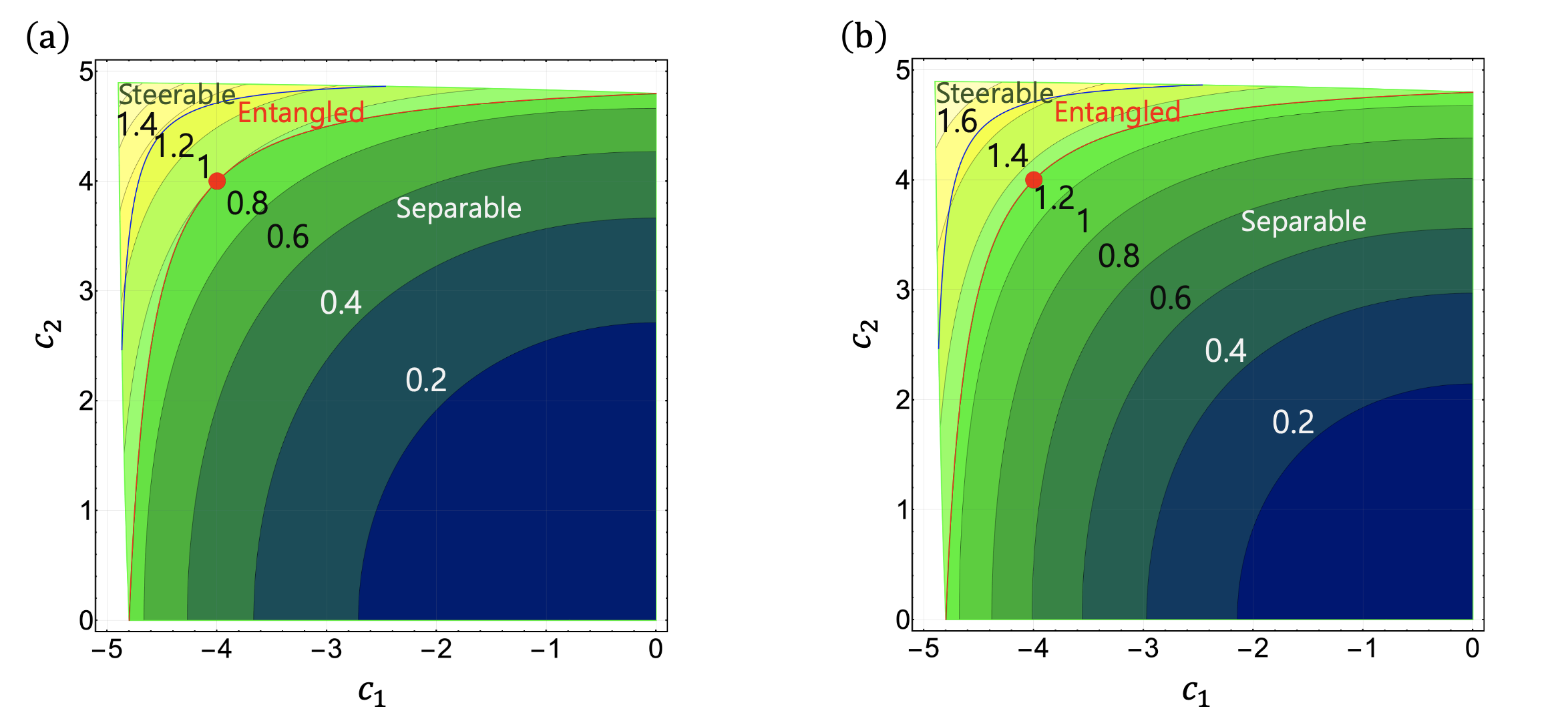}
    \caption{Contour plots of the extractable work as functions of $c_1$ and $c_2$, in the second quadrant, under (a) homodyne measurement and (b) heterodyne measurement for $a=b=5$. Different types of correlated states occupy distinct regions and the amount of extractable work increases with increasing correlation strengths, i.e., either $|c_1|$ or $|c_2|$. Red dots indicate where the maximum value of extractable work is obtained on the lines separating entangled states from separable states.}
    \label{result:fig2}
\end{figure}

Figure \ref{result:fig2} depicts $W_{\rm hom}$ and $W_{\rm het}$ as functions of $c_1$ and $c_2$ for $a=b=5$. Similar plots are obtained as the magnitudes of $a=b$ is varied.
Clearly, the amount of extractable work increases with the strengths of the correlations $|c_1|$ and $|c_2|$. For a fixed value of $|c_1|$, both $W_{\rm hom}$ and $W_{\rm het}$ increase with increasing $|c_2|$ and vice versa. As in the previous case in which $c_1 = -c_2$, different types of correlations cannot be distinguished using the work value alone. However, the hierarchy $W^{\rm separable}< W^{\rm entangled}<W^{\rm steerable}$ is maintained as the correlation strength increases. The heterodyne protocol extracts more work than the homodyne protocol at any given point in the parameter space, 

The red dot indicates where the maximum value of work is obtained on the line delineating separable and entangled states, which is placed exactly at $c_1 = -c_2$. As we see in the next subsection, this point moves to the edges if $a$ differs substantially from $b$. On the contrary, the maximum extractable work is always obtained at $c_1 = -c_2$ on both the entangled/steerable line and the steerable/non-physical line.

\subsection{$a\neq b$ and $c_1\neq c_2$ }
Lastly, we consider the most general case where all the parameters, $a$, $b$, $c_1$ and $c_2$, are independent. The phase diagram has the same symmetry as in the $a=b$ case, so we can concentrate on the second quadrant. In this case, separable states obey $c_2 \le \frac{c_1}{a b - c_1^2} + \sqrt{a b + \frac{a b}{(c_1^2-a b)^2} + \frac{a^2 + b^2}{c_1^2-a b}}$ while non-steerable states obey $ c_2 \le \sqrt{b \left( a + \frac{b}{c_1^2-a b} \right)}$.

\begin{figure} [!h]
    \centering
 \includegraphics[width=0.95\columnwidth]{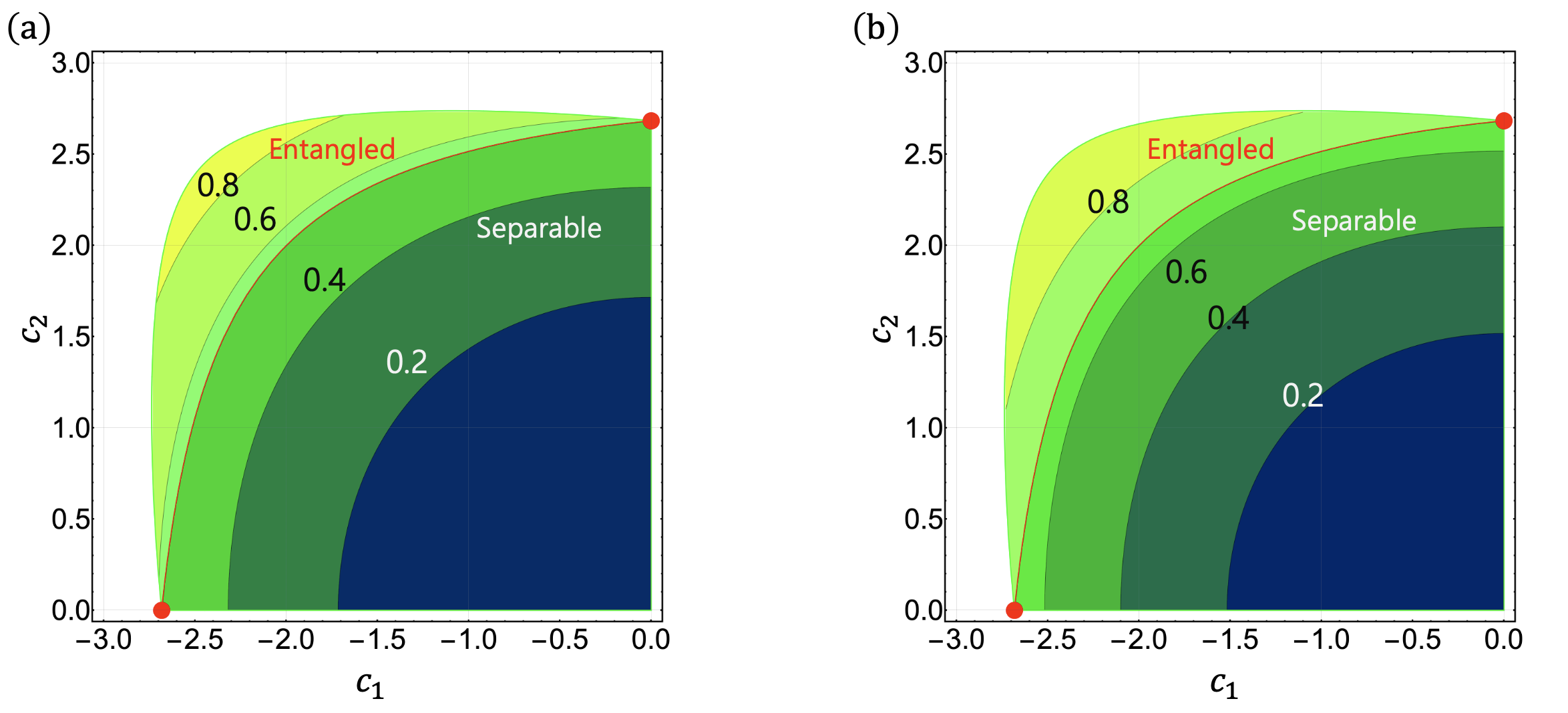}
    \caption{Contour plots of the extractable work as functions of $c_1$ and $c_2$, in the second quadrant, under (a) homodyne measurement and (b) heterodyne measurement for $a=5$ and $b=2$. Maximum value of extractable work for the separable states are obtained at either $c_1 =0$ or $c_2 =0$, as indicated by the red dots. }
    \label{result:fig3}
\end{figure}

As the local energy of the subsystem B is lowered while keeping the local energy of the subsystem A constant, i.e.~$a>b$, the steerable region shrinks and eventually disappears at $b = (1+a)/2$. No steerable states exist below this value of $b$. On the other hand, the red dot suddenly shifts to the edges of the quadrant--$c_1 = 0$ and $c_2 =0$--as shown in Fig.~\ref{result:fig3}. In the homodyne (heterodyne) case, this occurs at $b \approx 2.56 $ ($b = 3$) for $a=5$. For the heterodyne protocol, this point can be obtained analytically, which reads $b = (1+a)/(a-3)$. Apart from the lack of steerable states, we reach similar conclusions as in the previous two parameter regimes. Separable and entangled states occupy two distinct regions of the parameter space and the hierarchy $W^{\rm separable}< W^{\rm entangled}$ with increasing correlation strength remains intact.

\begin{figure}[!h]
    \centering
\includegraphics[width=0.95\columnwidth]{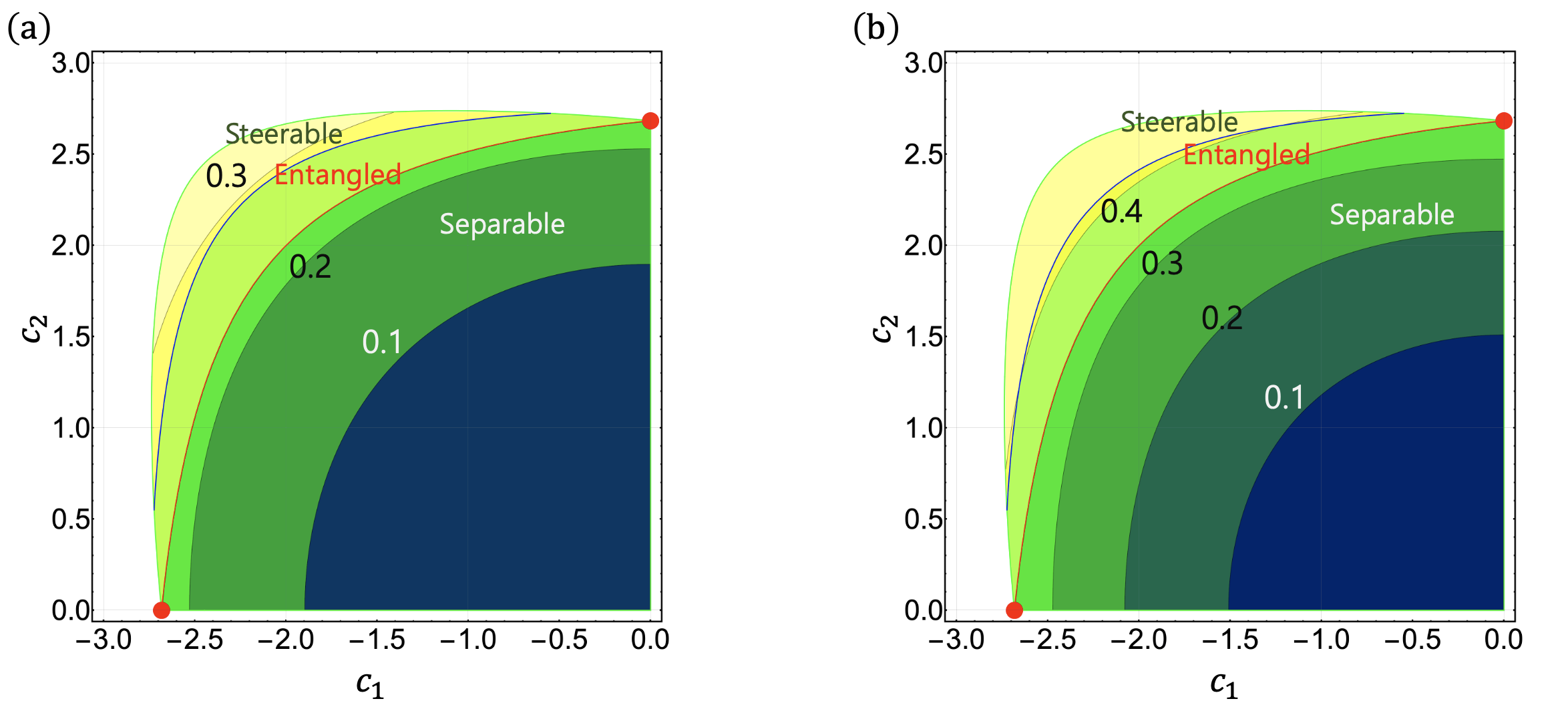}
    \caption{Contour plots of the extractable work as functions of $c_1$ and $c_2$, in the second quadrant, under (a) homodyne measurement and (b) heterodyne measurement for $a=2$ and $b=5$. Maximum value of extractable work for the separable states are obtained at either $c_1 =0$ or $c_2 =0$, as indicated by the red dots.}
    \label{result:fig4}
\end{figure}

We are left with one final parameter regime, namely $a<b$. A typical plot is displayed in Fig.~\ref{result:fig4}. The hierarchy remains intact, i.e., different types of correlated states occupy distinct regions of the parameter space. Moreover, the amount of extractable work is closely related to the correlation strength. We also note that there are always steerable states when $b>a$ and that the red dot suddenly jumps to the edges at $a \approx 2.56$ and $4$ for the homodyne and heterodyne protocols respectively (keeping $b=5)$.

\section{Conclusions}
In this work we have investigated the relation between extractable work and quantum correlation for a general two-mode Gaussian state, extending the previous study \cite{Mauro} to consider three different classes of quantum states, i.e., separable, entangled, and steerable states, and modifying the definition of work.  Specifically, we defined the work as the change in energy between the initial and final states of a local system, following local unitary actions conditioned on the outcome of a Gaussian measurement performed by the other party. Explicit protocols have been proposed for work extraction under homodyne and heterodyne measurements (performed by Bob), maximizing the extracted work through appropriate unitary Gaussian operations, i.e., local displacement and squeezing (performed by Alice). As a result, we have identified an overall hierarchy in the extracted work, $W^{\rm separable}< W^{\rm entangled}<W^{\rm steerable}$, depending on the strength of quantum correlation.  

As a remark, we have only considered the change in Alice's local energy resulting from the extraction of information by Bob's measurement on his local system. This led us to naturally explore Bob $\rightarrow$ Alice steering. If the measurement is performed by Alice instead, Alice $\rightarrow$ Bob steering should be considered. However, this scenario is already accounted for in our study, as interchanging  $a$ and $b$ reverses the roles of the two parties.

In the future, it would be an interesting yet nontrivial task to rigorously define a continuous variable quantum heat machine that manifests quantum steering through enhanced work extraction--a feat  achieved so far only in discrete-variable quantum heat machines \cite{Strunz,Ji}. In addition, extending this study to non-Gaussian states and measurements would be desirable \cite{nha0,nha1,nha2}, as it will further enhance our understanding of the role played by quantum correlations in the working principles of quantum thermodynamics.

\section*{Acknowledgements}

J.L. and C.N. are supported by RS-2023-NR119931 through the National Research Foundation of Korea (NRF) funded by the Korea government (MSIT). K.J. and H.N. are supported by NRF-2022M3H3A1098237 through the National Research Foundation of Korea (NRF) funded by the Korea government (MSIT).

\end{document}